\documentclass[11pt]{article}
\baselineskip
\usepackage{amsmath,amssymb}
\usepackage{graphicx}
\usepackage{cite}
\usepackage{tabls}
\usepackage{color}
\usepackage{hyperref}
\usepackage[english]{babel}
\usepackage{comment}

\newcommand{\SU}{\mathrm{SU}}
\newcommand{\xiratio}{\xi/\xi_{2nd}}
\newcommand{\Tr}{{\rm Tr\,}}

\begin{document}

\hypersetup{pageanchor=false}
\begin{titlepage}
%\maketitle
\begin{center}
{\Large\bf $\xiratio$ ratio as a tool to refine Effective Polyakov Loop models}
\end{center}
\vskip1.3cm
\centerline{Michele Caselle and Alessandro Nada}
\vskip1.5cm
\centerline{\sl Department of Physics, University of Turin \& INFN, Turin}
\centerline{\sl Via Pietro Giuria 1, I-10125 Turin, Italy}
\vskip0.5cm
\begin{center}
{\sl  E-mail:} \hskip 1mm \href{mailto:caselle@to.infn.it}{{\tt caselle@to.infn.it}}, \href{mailto:anada@to.infn.it}{{\tt anada@to.infn.it}}
\end{center}
\vskip1.0cm
\begin{abstract}
Effective Polyakov line actions are a powerful tool to study the finite temperature behaviour of lattice gauge theories.
They are much simpler to simulate than the original lattice model and are affected by a milder sign problem, but it is not clear to which extent they really capture the rich spectrum of the original theories.
We propose here a simple way to address this issue based on the so called second moment correlation length $\xi_{2nd}$.
The ratio $\xiratio$ between the exponential correlation length and the second moment one is equal to 1 if only a single mass is present in the spectrum, and it becomes larger and larger as the complexity of the spectrum increases.
Since both $\xi$ and $\xi_{2nd}$ are easy to measure on the lattice, this is a cheap and efficient way to keep track of the spectrum of the theory.
As an example of the information one can obtain with this tool we study the behaviour of $\xiratio$ in the confining phase of the ($D=3+1$) $\mathrm{SU}(2)$ gauge theory and show that it is compatible with 1 near the deconfinement transition, but it increases dramatically as the temperature decreases.
We also show that this increase can be well understood in the framework of an effective string description of the Polyakov loop correlator.
This non-trivial behaviour should be reproduced by the Polyakov loop effective action; thus, it represents a stringent and challenging test of existing proposals and it may be used to fine-tune the couplings and to identify the range of validity of the approximations involved in their construction.
\end{abstract}

\end{titlepage}
\hypersetup{pageanchor=true}

\section{Introduction}

The study of the phase diagram of quantum chromodynamics (QCD) from first principles on the lattice is severely limited by the well-known ``sign problem'': when the quark chemical potential $\mu$ is added to the theory, the fermion determinant becomes complex, making Monte~Carlo simulations unfeasible as importance sampling is not applicable anymore.
Many different methods have been proposed to circumvent this problem but while they are generally in agreement for small values of the chemical potential ($\mu \lesssim T$), reliable computations for larger values of $\mu$ are still missing.

In the last few years, several attempts to investigate the phase structure of QCD and QCD-like theories and to address the sign problem for any value of the chemical potential were instead based on a family of models called effective Polyakov Loop (EPL) models \cite{Gattringer:2011gq,Mercado:2012ue,Greensite:2012xv,Greensite:2012dy,Greensite:2013yd,Greensite:2013bya,Greensite:2014isa,Hollwieser:2016hne,Hollwieser:2016yjz,Langelage:2010yr,Fromm:2011qi,Bergner:2013qaa,Bergner:2015rza,Dittmann:2003qt,Heinzl:2005xv,Wozar:2006fi,Wozar:2007tz,Billo:1996wv,Aarts:2011zn,Scior:2014zga,Bahrampour:2016qgw}.
In a nutshell, the original theory regularized on the lattice is mapped to a three-dimensional, center-symmetric, effective Polyakov loop spin model obtained by integration over the gauge and matter degrees of freedom.
This type of approach is particularly interesting, since in the resulting effective description the sign problem is milder and can be treated with known methods, or it can be avoided entirely. 

An exact integration of timelike degrees of freedom is too difficult, but from strong-coupling expansions one can infer a few important features of such an effective action:
\begin{itemize}
 \item first, it should be nonlocal since, as the order of the strong-coupling expansion increases, far apart Polyakov loops are involved in the interaction;
 \item second, it should involve higher representations of the Polyakov loop;
 \item third, it should contain multispin interactions.
\end{itemize}

In most of the existing proposals multispin interactions are neglected, assuming that they can be taken into account by a suitable tuning of two-spin interactions.
As a result, one ends up with an action of this type:
\begin{equation}
S_{\text{eff}}=\sum_{p}\sum_{|{\bf r}|\geq 1}\sum_{|{\bf x}-{\bf y}|=\bf r} \lambda_{p,{\bf r}} \, \chi_p({\bf x}) \, \chi_p({\bf y})
\label{linansatz}
\end{equation}
where $\chi_p (\bf x)$ is the character in the $p$ representation of the loop in the spatial site $\bf x$ and $\lambda_{p,{\bf r}}$ is the coupling between the effective spins.

Several strategies have been devised to address the problem of determining the infinitely many interaction terms between the degrees of freedom in the new three-dimensional theory.
In general, the main idea of EPL proposals is to keep the number of free parameters in the action as small as possible and try to summarize in a few relevant couplings the complexity of the original model.

The purpose of this paper is to suggest a quantitative way to test these proposals, to evaluate their ability to capture the relevant features of the original four-dimensional model, and, possibly,
to fine-tune the coupling constants $\lambda_{p,{\bf r}}$.
More precisely, we propose the ratio between the exponential correlation length $\xi$ and the second moment correlation length $\xi_{2nd}$ as a tool to test existing EPL actions: indeed, this ratio is well defined for any model and it can be readily evaluated in Monte~Carlo simulations on the lattice.

This study has been inspired by the observation that any EPL proposal must face two nontrivial requirements, which are indeed two faces of the same coin.

The first issue is that Polyakov loop correlators extracted from EPL actions should display the so-called ``L\"uscher term'' \cite{Luscher:1980fr}, i.e. a $1/R$ correction in the static quark-antiquark potential.
Such a term is present in the confining phase of the original model and has been detected and precisely measured on the lattice in $\SU(N)$ gauge theories, both using Wilson loops \cite{Necco:2001xg} and Polyakov loop correlators \cite{Luscher:2004ib,Caselle:2004er,Athenodorou:2010cs,Athenodorou:2011rx}. 
EPL actions should be able to reproduce the same behavior. 
This is indeed a very nontrivial requirement, since such a term is typical of extended gauge invariant observables in gauge theories, and, in general, spin models with short distance interactions do not show such a behavior.

The second issue is that the original lattice gauge theory (LGT) is characterized by a rich spectrum of excitations, with an exponential ``Hagedorn'' type of dependence on the energy, which, again, is typical of gauge theories and it is not easy to reproduce with a spin model.

These two features are deeply related: it is exactly the accumulation of an infinite number of excitations which leads to the $1/R$ correction in the potential. 
This relation can be shown explicitly in the framework of the so-called effective string Description of LGTs using the open-closed string duality (or, equivalently, performing a modular transformation of the effective string result for the interquark potential).  
We shall address this issue in Sec.~\ref{sec:est} of this paper.

A simple and easy way to keep track of the spectrum of a statistical model is exactly the $\xi/\xi_{2nd}$ ratio that we discuss in this paper. 
Our idea is to use this easily computable quantity to understand if the spectrum is dominated by a single mass or contains several masses in competition among them. 
This information could help to select the terms to be included in the effective action and, possibly, to fine-tune the couplings obtained with the existing approaches.
We shall discuss as an example the case of the $(3+1)$-dimensional $\SU(2)$ Yang-Mills theory and compare the results with the simplest possible EPL for this theory, namely the nearest-neighbor Ising model in three dimensions. 

This paper is organized as follows: in Sec.~\ref{sec:xi2nd_ising} we shall introduce the second moment correlation length $\xi_{2nd}$ and discuss its relation with the exponential correlation length $\xi$. 
Section~\ref{sec:su2results} is devoted to the results of the simulation of the $\SU(2)$ theory and to a comparison with the three-dimensional Ising model. 
In Sec.~\ref{sec:est} we shall discuss these results in the framework of the effective string picture and finally, Sec.~\ref{sec:conclusion} will be devoted to some concluding remarks. 

\section{The relation between $\xi$ and $\xi_{2nd}$ in spin models}
\label{sec:xi2nd_ising}

In a $d$-dimensional spin model, the exponential correlation length $\xi$ describes the long distance behavior of the connected two point function and is defined as
\begin{equation}
\label{exp_corr_length}
 \frac{1}{\xi}=-\lim_{|\vec{n}|\to\infty}\frac{1}{|\vec{n}|} \log\langle s_{\vec{0}} s_{\vec{n}}\rangle_c
\end{equation}
where $s_{\vec{x}}$ denotes the spin $s$ in the position $\vec{x}$, and the connected correlator is defined as
\begin{equation}
\langle s_{\vec{m}} s_{\vec{n}} \rangle_c = 
\langle s_{\vec{m}} s_{\vec{n}} \rangle -  \langle s_{\vec{m}}  \rangle ^2.
\end{equation}
The square of the second moment correlation length $\xi_{2nd}$ is defined as:
\begin{equation}
\xi_{2nd}^2 = \frac{\mu_2}{2 d \mu_0} \;,
\label{xi2nd}
\end{equation}
where 
\begin{equation}
\mu_0 = \lim_{L \rightarrow \infty}\; \frac{1}{V} \; \sum_{{\vec{m}},{\vec{n}}} \;
\langle s_{\vec{m}} s_{\vec{n}} \rangle_c
\end{equation}
and 
\begin{equation}
\mu_2 = \lim_{L \rightarrow \infty}\; \frac{1}{V} \;  \sum_{{\vec{m}},{\vec{n}}} \; |\vec{m} - \vec{n}|^2 \langle s_{\vec{m}} \, s_{\vec{n}} \rangle_c \;,
\end{equation}
where $V= L^d$ is the lattice volume, and the sum on $\vec{m}$ is on the $d$ indices $m_0,m_1,...,m_{d-1}$.

\par
This estimator for the correlation length was very popular a few years ago, since its numerical evaluation in Monte~Carlo simulations is simpler and faster than that of the exponential correlation length. 
Moreover it is the length scale which is directly observed in scattering experiments. 

It is important to notice that $\xi_{2nd}$ is not exactly equivalent to $\xi$. The difference is in general very small, but it carries important information on the spectrum of the underlying theory.
To better understand this issue it is useful to move to ``time-slice'' variables.
Let us fix for simplicity $d=3$, and let us denote $\vec{n}=(n_0,n_1,n_2)$, where $n_0$ denotes the ``time'' direction. 
We can define the time-slice variables as
\begin{equation}
 S_{n_0} = \frac{1}{L^2}\sum_{n_1,n_2} s_{(n_0,n_1,n_2)}
\label{timeslice}
\end{equation}
and the time-slice correlation function as 
\begin{equation}
G(\tau) = \sum_{n_0} \left\{ \langle S_{n_0} S_{{n_0}+\tau} \rangle - \langle S_{n_0} \rangle^2 \right\} \; ;
\label{correlator1}
\end{equation}
the wall average on the $(n_1,n_2)$ plane indicates a projection to zero spatial momentum.
As for the standard correlator, the exponential correlation length can be extracted from the large distance behavior of $G(\tau)$,
\begin{equation}
G(\tau) \sim\; \exp(-\tau/\xi) \; .
\end{equation}
The relation between $\xi$ and $\xi_{2nd}$ can be obtained by noticing that we can rewrite $\mu_2$ as follows:
\begin{eqnarray}
 \mu_2 &=& \frac{1}{V} \; \sum_{ \vec{m},\vec{n} } \; |\vec{n} - \vec{m}|^2 \;\; \langle s_{\vec{m}} \, s_{\vec{n}} \rangle_c \nonumber \\
       &=& \frac{1}{V} \; \sum_{ \vec{m},\vec{n} } \; \sum_{i=0}^{d-1} \;\; (n_i - m_i)^2 \;\;  \langle s_{\vec{m}} \, s_{\vec{n}} \rangle_c \nonumber \\
       &=& \frac{d}{V} \; \sum_{ \vec{m},\vec{n} } \; (n_0 - m_0)^2  \;\; \langle s_{\vec{m}} \, s_{\vec{n}} \rangle_c
\end{eqnarray}
where the last identity holds when the lattice is symmetric in all the $d$ directions, so that inside the sum over $\vec{m}$ and $\vec{n}$ the $d$ terms of the sum over $i$ are equal.

Because of the exponential decay of the correlation function, this sum is convergent, and we can commute the spatial summation with the summation over configurations so as to obtain
\begin{equation}
\mu_2 = d L^2 \sum_{\tau=-\infty}^{\infty} \; \tau^2 \;\; \langle S_0 \; S_\tau \rangle_c\;\;\; 
\end{equation}
and
\begin{equation}
\mu_0 = L^2 \sum_{\tau=-\infty}^{\infty} \; \langle S_0 \; S_\tau  \rangle_c \;\;\; 
\end{equation}
for which we used the definition given in Eq.~(\ref{timeslice}).
If we now insert these results in Eq.~(\ref{xi2nd}), we obtain
\begin{equation}
\xi_{2nd}^2=\frac{ \sum_{\tau=-\infty}^{\infty} \; \tau^2 \;\;
G(\tau)}{ 2 \sum_{\tau=-\infty}^{\infty} \;\;
G(\tau)}\;\;\; .
\label{eq13}
\end{equation}
Assuming a multiple exponential decay for $G(\tau)$,
\begin{equation}
\langle S_0 \; S_\tau  
\rangle_c \propto \sum_i \; c_i \; \exp(-|\tau|/\xi_i) \;\; , 
\label{3mass}
\end{equation}
and replacing the summation by an integration over $\tau$, we get
\begin{equation}
\label{e24}
 \xi_{2nd}^2 = \frac12 \;
          \frac{\int_{\tau=0}^{\infty} 
{\mbox d}\tau \;\tau^2 \sum_i c_i
\; \exp(-\tau/\xi_i)}
                     {\int_{\tau=0}^{\infty} {\mbox d}\tau \; \sum_i c_i
\exp(-\tau/\xi_i)}
           =   \frac{ \sum_i c_i \xi_i^3}{\sum_i c_i \xi_i} \;,
\end{equation}
which is equal to $\xi^2$ if only one state contributes. 
It is, thus, clear that we can use the $\xi/\xi_{2nd}$ to gain some insight on the spectrum of the theory and on the amplitude $c_i$ of these states.

In the following, we examine a set of results for the Ising universality class that is reported in Table~\ref{tab:xi_ising}: this class of models has not only been studied in great detail and with high accuracy, but it is also the relevant one for the test for the $\SU(2)$ gauge theory that we will discuss in Sec.~\ref{sec:su2results}.
The $d=2$ results are obtained from the exact solution of the two-dimensional Ising model, while the $d=3$ results are obtained from $\epsilon$ expansion calculations, Monte~Carlo simulations, or strong-coupling expansions.  
For a review of these and other results see, for instance, Ref.~\cite{Pelissetto:2000ek}.

\begin{table}[!htb]
\centering
%\centering
%\begin{ruledtabular}
\begin{tabular}{|l|c|c|l|}
\hline
 & $d$ & $\xiratio$ & Method\\
\hline
\hline
High-$T$ phase & $2$ & $1.00040...$  & \\
               & $3$ & $1.00016(2)$  & Strong-coupling + $\epsilon$ exp.~\cite{Campostrini:1999at}\\
               & $3$ & $1.00021(3)$  & Perturbative calc.~\cite{Campostrini:1997sn}\\
               & $3$ & $1.000200(3)$ & Strong-coupling~\mbox{\cite{Campostrini:1999at}}\\
\hline
\hline
Low-$T$ phase & $2$ & $1.58188...$ & \\
              & $3$ & $1.031(6)$   & MC simulations~\cite{Caselle:1999tm}\\
              & $3$ & $1.032(4)$   & Strong-coupling~\mbox{\cite{Campostrini:1999at}}\\
\hline
\hline
Critical isotherm & $2$ & $1.07868...$ & \\
($t = 0$,  $|H|\not= 0$) & $3$ & $1.024(4)$   & Strong-coupling + $\epsilon$ exp.~\mbox{\cite{Campostrini:1999at}}\\
\hline
\end{tabular}
\caption{Values of the $\xiratio$ ratio for an Ising spin system in three different conditions: in the high-temperature symmetric phase, in the low-temperature broken symmetry phase, and along the critical isotherm. 
It is important to notice that in $d=3$, there is a general agreement among the results obtained with very different approaches ranging from Monte~Carlo simulations to strong-coupling expansions. 
\label{tab:xi_ising}}
%\end{ruledtabular}
\end{table}

In the high-$T$ symmetric phase, where the spectrum is composed by a single massive state, we would expect that $\xi/\xi_{2nd}=1$: the small but not negligible difference from $1$ can be better understood looking at the Fourier transform of $G(\tau)$.
Besides isolated poles, which correspond to isolated exponentials in $G(\tau)$, we also have cuts above the pair production threshold at momentum $p$ equal to twice the lowest mass. 
In the original correlator, these cuts can be thought of as the coalescence of infinitely nearby exponentials\footnote{As a matter of fact on a finite lattice, this is their correct description, since the transfer matrix has only a finite number of eigenvalues.}, and as such, they also contribute to the ratio $\xi/\xi_{2nd}$.

In $d=3$, both in the low-$T$ broken symmetry phase and on the critical isotherm curve ($T=T_c$, $H\not=0$), the $\xi/\xi_{2nd}$ ratio is definitely larger: indeed, besides the cuts discussed above, we also have one (or more) isolated bound states which contribute to the spectrum. 
This is the case, for instance, of the 3d Ising model for $T<T_c$, for which an infinite tower of bound states exists \cite{Caselle:2001im}: in particular, the lowest of such states takes the value $m_{bound}=1.83(3) \, m_{ph}$ \cite{Caselle:1999tm} and is, thus, located below the two-particle threshold. 

The $d=2$, $T=T_c$, $H\not=0$ case is of particular interest: thanks to the exact solution of Zamolodchikov \cite{Zamolodchikov:1989fp} we know that there are three particles in the spectrum below the two-particle threshold, and accordingly, the difference $\frac{\xi}{\xi_{2nd}}-1$ is about 3 times larger than in the $d=3$, $T<T_c$ case, which, as we have seen, has only one state below threshold.

Finally, it is very instructive to look at the $d=2$, low-$T$ case, in which the Fourier transform of the correlators starts with a cut. 
Following the analysis of McCoy and Wu \cite{McCoy:1978ta} (and, more recently, of Fonseca and Zamolodchikov \cite{Fonseca:2001dc}), we know that we may interpret the spectrum of this model as the coalescence of an infinite number of states. 
Accordingly, a much larger value of the ratio $\xi/\xi_{2nd}$ is found, with a difference from $1$ which is 1 order of magnitude larger than the value in the presence of an isolated bound state and 3 orders of magnitude larger than the $T>T_c$ case.

\section{Numerical results for the ($D=3+1$)  $\SU(2)$ lattice gauge theory}
\label{sec:su2results}

We performed a numerical study of the $\xiratio$ ratio in the context of the $\SU(2)$ non-Abelian gauge theory, which is regularized on a finite hypercubic lattice of spacing $a$ and spacetime volume $\mathcal{V}=a^4 N_t \times N_s^3$.

The Yang-Mills action is discretized with the usual Wilson action~\cite{Wilson:1974sk}
\begin{equation}
%\label{wilson_action}
S_W = -\frac{2}{g^2} \sum_{x} \sum_{0 \le \mu < \nu \le 3} \Tr U_{\mu\nu} (x)
\end{equation}
where $U_{\mu\nu} (x)$ is the plaquette associated to the site $x$ and to the oriented plane in the $(\mu,\nu)$ directions, while $g^2$ is the bare coupling; we also introduce the inverse coupling $\beta=4/g^2$.

The main observable of interest is the so-called Polyakov loop, defined as the trace of a Wilson line winding around the lattice in the temporal (``$0$'') direction:
\begin{equation}
P(x,y,z) = \frac{1}{2} \Tr \prod_{0 \le t < N_t} U_0 (x,y,z,t \, a) \;.
\end{equation}
Furthermore, following the definition used for the spin model of Eq.~\ref{timeslice}, we can define the zero-momentum projection of the Polyakov loop by taking the average over two spatial dimensions
\begin{equation}
\label{zeromom}
\bar{P}(z) = \frac{1}{N_x N_y} \sum_{x,y} P (x,y,z)
\end{equation}
and we can write down the zero-momentum correlator $G(\tau)$,
\begin{equation}
 G(\tau) = \langle \bar{P}(0) \bar{P}(\tau) \rangle - \vert \langle P \rangle \vert^2
\end{equation}
which we can identify with the definition given in Eq.~(\ref{correlator1}).

The temperature $T$ is defined by the inverse of the shortest compactified dimension (which is usually identified as the temporal one) via the relation $T=1/a N_t$; the sizes of the ``spatial'' directions are chosen to be much larger ($N_s \gg N_t$), and periodic boundary conditions are imposed on all directions.
In practice, in order to change the temperature, both the number of sites $N_t$ and the lattice spacing $a$ can be varied. 
The lattice spacing in particular is controlled by the inverse bare coupling $\beta$: for the details concerning the determination of the relation between $a$ and $\beta$, we refer to Ref.~\cite{Caselle:2015tza}, in which the scale has been set using the square root of the string tension $\sigma$. 
In order to express our results in terms of the critical temperature, we used the value $T/T_c=0.7091(36)$ computed in Ref.~\cite{Lucini:2003zr}.
The setup of the numerical simulations, including the corresponding temperatures, is reported in Table~\ref{tab:lattice_setup}.

\begin{table}[!htb]
\centering
%\begin{ruledtabular}
\begin{tabular}{|c|c|c|c|}
\hline
$\beta$ & $ N_s^3 \times N_t$ & $T/T_c$ & $n_{conf}$\\
\hline
\hline
$2.27$  & $32^3 \times 6$ & $0.59$ & $4.5 \times 10^5$ \\
$2.33$  & $32^3 \times 6$ & $0.71$ & $2.25 \times 10^5$ \\
$2.3$   & $32^3 \times 5$ & $0.78$ & $5.5 \times 10^5$ \\
$2.357$ & $32^3 \times 6$ & $0.78$ & $2.25 \times 10^5$ \\
$2.25$  & $64^3 \times 4$ & $0.84$ & $3 \times 10^4$ \\
$2.4$   & $64^3 \times 6$ & $0.90$ & $2 \times 10^4$ \\
\hline
\end{tabular}
\caption{Setup of the lattice simulations performed for the $\SU(2)$ gauge theory in the confined phase. In the first column we report the inverse coupling $\beta$ and in the second the spatial and temporal sizes, while the resulting temperature in units of $T_c$ and the statistics for the measurements of the zero-momentum Polyakov loop correlator are shown in the third and fourth columns, respectively.
\label{tab:lattice_setup}}
%\end{ruledtabular}
\end{table}

The data for the zero-momentum correlator have been fitted with a functional form of the type
\begin{equation}
\label{exp_corr_length_2}
 G(\tau) \sim \exp\left(-\frac{\tau}{\xi}\right) + \exp\left(-\frac{L-\tau}{\xi}\right)
\end{equation}
where $\xi$ is the estimate of the exponential correlation length of the system as defined in Eq.~(\ref{exp_corr_length}), and $L$ is the size of the three-dimensional space (in our case, $L=N_s$).
Numerical results for $\xi$ in units of the lattice spacing at different values of the temperature $T$ are reported in Table~\ref{tab:xi_xi2nd}.
For each of the fits at different values of $\beta$, the first two values of the distance $\tau$ have been excluded in order to avoid lattice discretization artifacts.

The results of the fit for $\xi$ have been compared also with an effective definition of the correlation length given by
\begin{equation}
 \xi_{eff}(\tau) = \frac{1}{\log(G(\tau+1))-\log(G(\tau))}.
\end{equation}
The $\xi_{eff}(\tau)$ stabilizes to values which are in good agreement with the one extracted using Eq.~(\ref{exp_corr_length_2}).

For the simulations close to the deconfinement transition (those reported in line 5 and line 6 of Table~\ref{tab:xi_xi2nd}), the spatial size $L$ of the lattice was increased in order to accommodate for a larger correlation length and to reduce the effect of the echo due to the periodic boundary conditions imposed on the spatial directions.

Next, we computed $\mu_2$ and $\mu_0$: in order to do so, we followed a procedure similar to that used in Ref.~\cite{Caselle1997a}. For $\mu_2$ we used an ansatz of the type
\begin{equation}
\label{mu2_calc}
 \mu_2 = \sum_{\tau=1}^{\tau_{max}} \tau^2 \, G(\tau) + \sum_{\tau=\tau_{max}+1}^\infty \tau^2 \, G(\tau_{max}) \, e^{-(\tau-\tau_{max})/\xi }
\end{equation}
and similar for $\mu_0$. 
The contribution of the tail, i.e., of the second term of Eq.~\eqref{mu2_calc}, depends on the value assigned to the distance cutoff $\tau_{max}$: during our analysis. we kept $\tau_{max} \sim 3 \xi$, which generally yielded stable results. 
Using Eq.~\eqref{xi2nd} (with $d=3$), we computed $\xi_{2nd}$ for different values of the temperature $T$: the results are reported in Table~\ref{tab:xi_xi2nd}.

\begin{table}[!ht]
\centering
%\begin{ruledtabular}
\begin{tabular}{|c|c|c|c|c|}
\hline
$T/T_c$ & $N_s$ & $\xi/a$ & $\xi_{2nd}/a$ & $\xi/\xi_{2nd}$\\
\hline
\hline
$0.59$ & $32$ & $1.31(2)$ & $0.887(8)$  & $1.48(3)$ \\
$0.71$ & $32$ & $2.31(4)$ & $1.842(15)$ & $1.25(2)$ \\
$0.78$ & $32$ & $2.56(2)$ & $2.22(1)$   & $1.153(11)$ \\
$0.78$ & $32$ & $3.08(4)$ & $2.67(2)$   & $1.151(16)$ \\
$0.84$ & $64$ & $3.05(6)$ & $2.74(4)$   & $1.11(3)$ \\
$0.90$ & $64$ & $6.9(2)$  & $6.6(3)$    & $1.04(6)$  \\
\hline
\end{tabular}
\caption{Results for the exponential correlation length $\xi$ (third column) and the second moment correlation length $\xi_{2nd}$ (fourth column) in units of the lattice spacing, along with their ratio $\frac{\xi}{\xi_{2nd}}$ (fifth column) in the confined phase: the temperature $T=1/(a(\beta)N_t)$ (first column) is varied using the inverse coupling $\beta$ and the temporal extent $N_t$. 
\label{tab:xi_xi2nd}}
%\end{ruledtabular}

\end{table}

\begin{figure}[ht!]
\begin{center}
\includegraphics*[width=\textwidth]{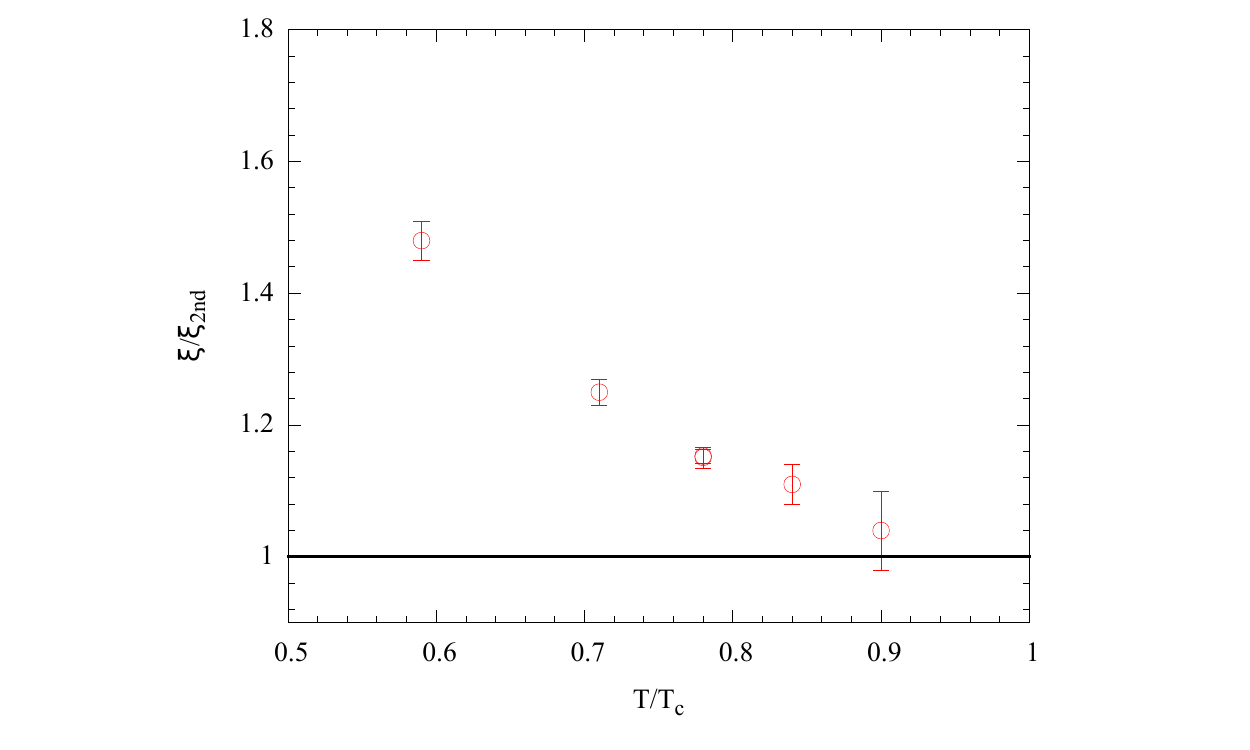}%
\caption{The $\xiratio$ ratio for different values of the temperature $T/T_c$ in the confining region.
\label{fig:xi2nd}
}
\end{center}
\end{figure}

The somewhat arbitrary choice of $\tau_{max}$ introduces a systematic error in the computation of $\mu_0$ and $\mu_2$: in the former case, such error is negligible since the contribution of the tail is very small compared to the first term. 
On the contrary, $\mu_2$ has a substantial contribution coming from large values of $\tau$, especially for temperatures close to $T_c$; however, we checked that the results obtained by changing $\tau_{max}$ were all within statistical error with each other. 

The data show a clear dependence on $T/T_c$ and in particular a dramatic increase of $\xiratio$ as the temperature decreases. In order to test if there are other dependences in our results, we realized the $T/T_c=0.78$ case with two different combinations of $\beta$ and $N_t$. We found the same value of $\xiratio$ even if the values of $\xi$ and $\xi_{2nd}$ were quite different in the two cases: this result makes us confident that scaling corrections are under control and that our results are tracing a true physical behavior of the $\SU(2)$ model.

\subsection{Comparison with the 3d Ising model.}
\label{sec:ising_comparison}

The simplest possible EPL model for the $\SU(2)$ lattice gauge theory discussed in the previous section is the 3d Ising model, which corresponds to the case in which in Eq.~(\ref{linansatz}) we truncate the action to the nearest-neighbour term, choose only the fundamental representation, and approximate the Polyakov loop with its sign. 
This is a very crude truncation, but in the $\SU(2)$ case, in the vicinity of the deconfinement transition, it turns out to be a very good approximation. 
In fact, the Svetitsky-Yaffe conjecture~\cite{Svetitsky:1982gs} tells us that if a $(d+1)$ LGT with gauge group $G$ has a second order deconfinement transition and the $d$-dimensional spin model with a global symmetry group being the center of $G$ has a second order magnetization transition, then the two models belong to the same universality class. 
This is exactly the case of the $(D=3+1)$ $\SU(2)$ LGT and of the 3d Ising model. It is easy to see using symmetry arguments that the Polyakov loop (order parameter of the deconfinement transition) is mapped by this identification into the spin of the Ising model (which is, in fact, the order parameter of the magnetization transition) and that the confining phase (the one which we studied in the previous section) is mapped into the high-temperature symmetric phase of the Ising model.
From the discussion of Sec.~\ref{sec:xi2nd_ising}, we, thus, expect the $\xiratio$ ratio to be very close to $1$ when the $\SU(2)$ model is close to the deconfinement transition, as this is the observed behavior in the high-$T$ phase for the 3d Ising model (as reported in Table~\ref{tab:xi_ising}).
This is indeed the case for the highest value of $T/T_c$ that we simulated. 
However, we see from the data that the ratio increases dramatically as $T/T_c$ decreases. 
This shows that as $T/T_c$ decreases, the Ising approximation becomes indeed too crude, and more sophisticated EPL models are needed. 
We will see below that this increase in the $\xiratio$ is essentially due to the combination of two nontrivial features of the gauge theory spectrum: 
\begin{itemize}
\item
first, the fact that as $T/T_c$ decreases, the states of the spectrum coalesce toward the ground state, exactly as it happens in the $d=2$ Ising model below $T_c$;
\item
second, the fact that the overlap constants $c_i$ increase exponentially with the energy of the states.
\end{itemize}
The nearest-neighbour Ising model cannot mimic these two features and, thus, must be discarded as $T/T_c$ decreases. 
It is interesting to notice as a side remark that our analysis offers a nice and simple way to estimate the range of validity of the the Svetitsky-Yaffe conjecture which, within our range of precision, holds for $T/T_c>0.9$.

\section{Effective string description of the interquark potential}
\label{sec:est}

A very useful tool to understand both the features of the spectrum mentioned above is the effective string description of the Polyakov loop correlators~\cite{Luscher:1980ac,Luscher:1980fr}, which indeed predicts, as a consequence of the ``string'' nature of the color flux tube, a rich spectrum of excitations. 
In particular, it has been recently realized that the Nambu-Got\={o} action~\cite{Nambu:1974zg,Goto:1971ce} is a very good approximation of this effective string model~\cite{Luscher:2004ib,Aharony:2013ipa}. 
The Nambu-Got\={o} action is simple enough to be exactly solvable so that the spectrum can be studied explicitly: for example, the large distance expansion of the Polyakov loop correlator in $D$ spacetime dimensions is \cite{Luscher:2004ib, Billo:2005iv}
\begin{eqnarray}
  \left\langle P(x)^{\ast}\kern-1pt P(y)\right\rangle  = & \sum_{n=0}^{\infty}w_n\frac{2r \sigma N_t}{{E}_n} \left(\frac{\pi}{\sigma}\right)^{\frac{1}{2}(D-2)} \nonumber\\
  &\times \left(\frac{{E}_n}{2\pi r}\right)^{\frac{1}{2}(D-1)} K_{\frac{1}{2}(D-3)}({E}_nr)
\label{NG}
\end{eqnarray}
where $w_n$ denotes the multiplicity of the state, $N_t$ the size of the lattice in the compactified time direction, and $E_n$ the closed-string energies which are given by
\begin{equation}
  {E}_n=\sigma N_t
  \left\{1+\frac{8\pi}{\sigma N_t^2}\left[-\frac{1}{24}\left(D-2\right)+n\right]
  \right\}^{1/2}.
\end{equation}
The weights $w_n$ can be easily obtained from the expansion in series of $q$ of the infinite products contained in the Dedekind functions which describe the large-$r$ limit of Eq.~(\ref{NG}) (see Ref.~\cite{Billo:2005iv} for a detailed derivation):
\begin{equation}
\label{etaexp}
\left(\prod_{r=1}^\infty \frac{1}{1 - q^r}\right)^{D-2}
= \sum_{k=0}^\infty w_k q^k.
\end{equation}
For $D=3$, we have simply $w_k=p_k$, the number of partitions of the integer $k$, while for $D>3$, these weights can be straightforwardly obtained from combinations of the $p_k$.
It is important to notice that these weights diverge exponentially as $n$ increases; in particular, we have
\begin{equation}
w_n \sim \exp\left({\pi\sqrt{\frac{2(D-2)n}{3}}} \right)\;.
\label{eq26}
\end{equation}
In the $D=3+1$ case we are interested in, thanks to the identity
\begin{equation}
 K_{\frac{1}{2}}(z)=\sqrt{\frac{\pi}{2z}}e^{-z}
\nonumber
\end{equation}
Eq.~(\ref{NG}) becomes
\begin{equation}
  \left\langle P(x)^{\ast}\kern-1pt P(y)\right\rangle = \sum_{n=0}^{\infty}\frac{N_t}{2r} w_n e^{-{E}_nr} \;
\end{equation}
which represents a collection of free particles of mass ${E}_n$.

\begin{figure}[ht!]
\begin{center}
\includegraphics[width=\textwidth]{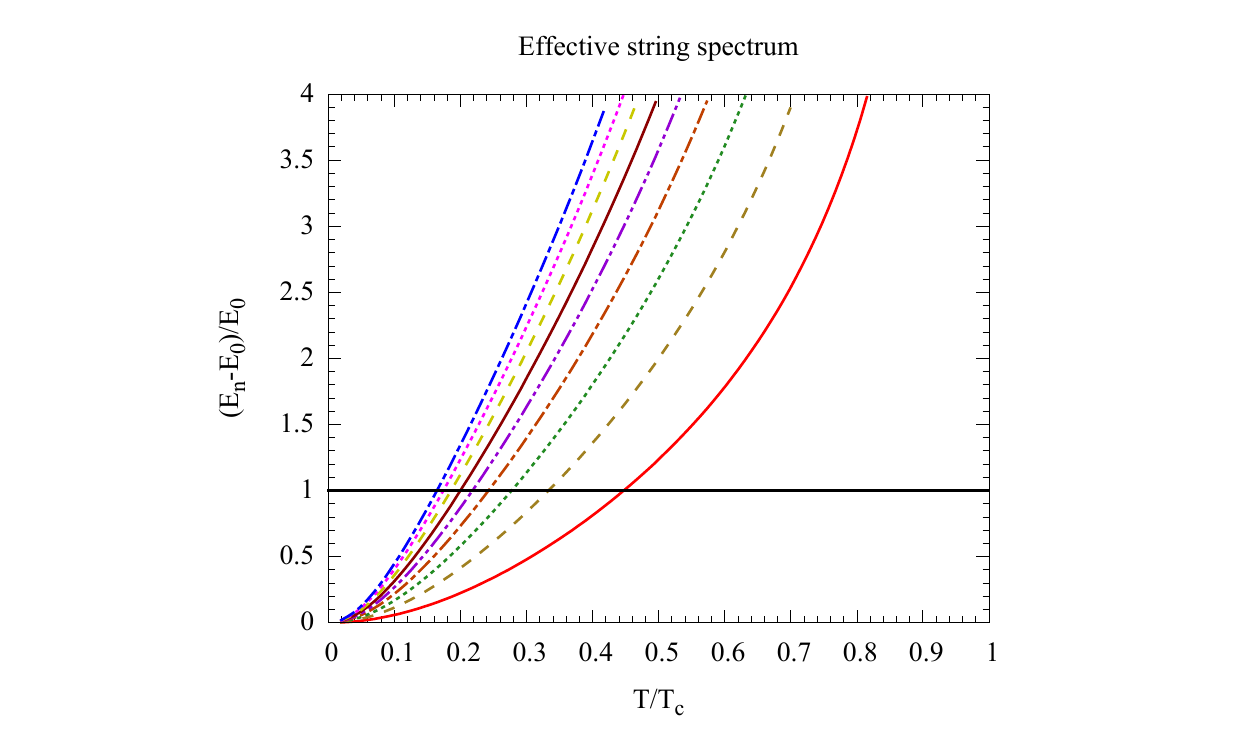}
\caption{The $(E_n-E_0)/E_0$ ratio as a function of $T/T_c$ for the first nine states, from $n=1$ (red solid line) to $n=9$ (blue dashed-dotted line). The black horizontal line represents the two-particle threshold.
\label{fig:masses}
}
\end{center}
\end{figure}

\begin{figure}[ht!]
\begin{center}
\includegraphics[width=\textwidth]{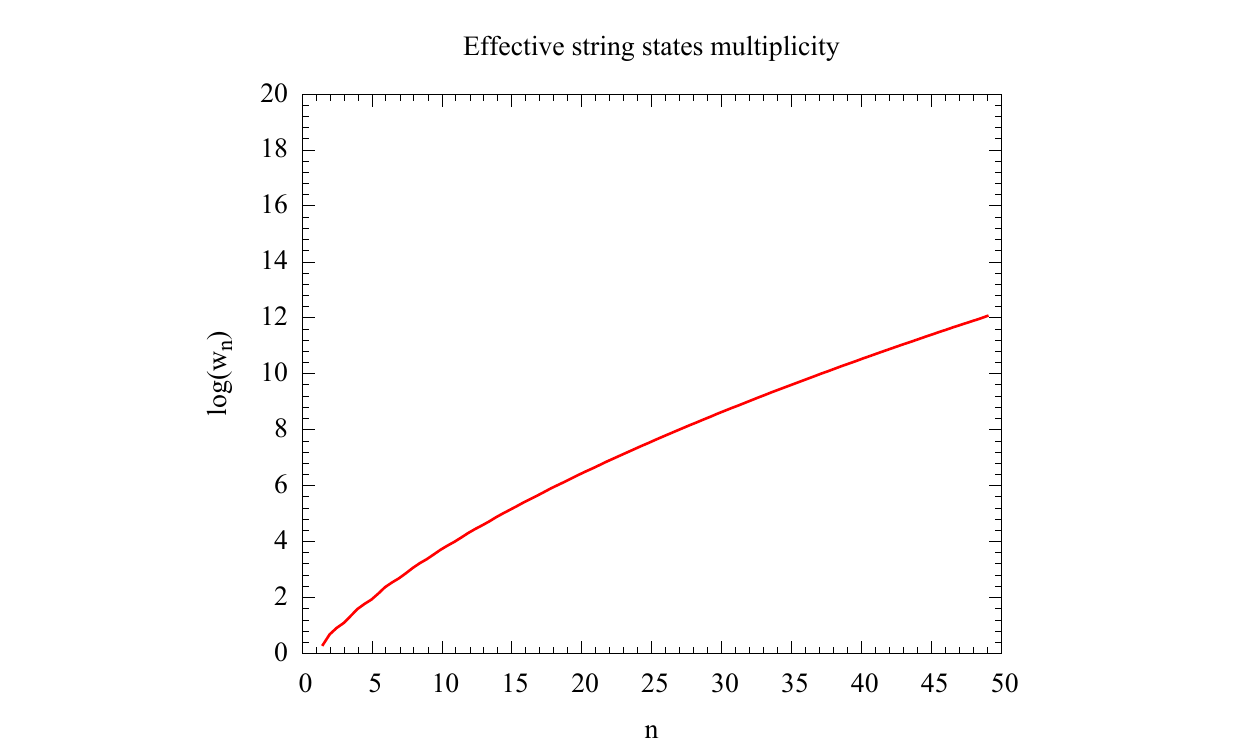}
\caption{The state multiplicity $w_n$ defined in Eq.~(\ref{NG}) as a function of the level $n$ for the first 50 levels.
\label{fig:multiplicity}
} 
\end{center}
\end{figure}

In the framework of the Nambu-Got\={o} approximation, one can also derive an estimate of the critical temperature $T_c$  measured in units of the square root of the string tension $\sqrt{\sigma}$ \cite{Olesen:1985e}
\begin{equation}
\frac{T_c}{\sqrt{\sigma}}=\sqrt{\frac{3}{\pi(D-2)}}
\end{equation}
given by the value of the ratio $\frac{T_c}{\sqrt{\sigma}}$ for which the lowest mass $E_0$ vanishes.
We can, thus, rewrite the energy levels as a function of $T/T_c$; setting $D=3+1$, we find
\begin{equation}
  {E}_n= \frac{2\pi T_c^2}{3 T} \left\{ 1 + 12 \, \frac{T^2}{T^2_c}\left[n-\frac{1}{12}\right] \right\}^{1/2}.
\end{equation}
This equation gives us a concrete realization of the two nontrivial features of the spectrum that we mentioned above.
\begin{itemize}
\item First, the gap between the different states {\sl decreases} as $T/T_c$ decreases, and all the states tend to accumulate toward the lowest state.
This is clearly visible in Fig.~\ref{fig:masses}, where we plotted the differences $(E_n-E_0)/E_0$ as a function of $T/T_c$ for the first ten states. 
The horizontal line represents the two-particle threshold.
\item
Second, the amplitudes $w_n$ increase exponentially with $n$. 
We plot the first 50 terms of this expansion in Fig.~\ref{fig:multiplicity}.
\end{itemize}

The combination of these two effects drives the $\xi/\xi_{2nd}$ ratio to larger values as the temperature decreases, as represented in Fig.~\ref{fig:xi2nd}. 
It is this behavior which EPL actions should be able to mimic, and in our opinion, it represents a stringent test for existing proposals.

Notice, as a side remark, that it is exactly the accumulation of infinite massive excitations toward the lowest state which drives the ``L\"uscher'' $1/R$ term in the low-$T$ regime of the theory. 
It is unlikely that such a term could be obtained by any mechanism other than an accumulation of infinite poles, in a local 3d spin theory like the existing proposal for EPL actions. 
It is also interesting to observe that the values that we measure of the  $\xi/\xi_{2nd}$ ratio in the $\SU(2)$ lattice gauge theory are significantly larger than the single state expectation even for temperatures in the range $0.6<T/T_c<0.9$, where (see Fig.~\ref{fig:masses}) there are no masses below the two-particle threshold. 
This large value is due to the exponential increase of the weights of the excited states with their energy as shown in Eq.~(\ref{eq26}). 
This is a typical ``stringlike'' effect, and it is exactly this type of signatures that the EPL model should be able to reproduce.

Unfortunately, it is not easy to obtain quantitative estimates of the $\xi/\xi_{2nd}$ ratio within the effective string framework. The reason lies in the fact that this description holds only for distances larger than a critical scale $r_c$, which in the Nambu-Got\={o} case is fixed to be $r_c=1/\sqrt{2\sigma}$ from the exponential divergence of the weights $w_k$, while it can be shown that a significant contribution to the ratio comes exactly from the distances below $r_c$. 
On the other side this makes the $\xi/\xi_{2nd}$ ratio a perfect tool to explore the small distance completion of the effective string description of the interquark potential: this is an issue that we plan to address in a future publication.

\section{Concluding remarks}
\label{sec:conclusion}

In this work, we introduced a novel tool for the refinement and fine-tuning of effective Polyakov line models in the form of the ratio between the exponential correlation length $\xi$ and the second moment correlation length $\xi_{2nd}$.
In Sec.~\ref{sec:su2results} we performed a numerical analysis in the context of the $\SU(2)$ lattice pure gauge theory, which showed how the original LGT at low temperatures presents values of the $\xiratio$ ratio that are much larger than 1. 
This feature, which is easy to study with lattice simulations, indicates that the theory possesses an extremely rich spectrum which must be taken into account when using EPL models to probe this region of temperature. 
In principle, one could certainly reproduce the correct spectrum of the theory by a careful matching of the correlator of the original theory with that of the EPL model. 
However, this may be a quite expensive strategy when the space of possible terms in the effective action is large (say, if one must decide if higher representation terms should be taken into account or if large distance couplings in the effective action must be truncated). 
In this respect, the $\xi/\xi_{2nd}$ ratio represents an economic and easily computable tool to perform this preliminary selection. 
Moreover, in order to understand the mechanisms underlying the behavior of the spectrum, we presented in Sec.~\ref{sec:est} the prediction of the effective string theory for the Polyakov loop correlator: this approach provides a clearer picture of the features of the spectrum, in particular, in terms of the coalescence of the states and the increase in their multiplicity.

It is interesting to notice that there is a natural implementation in effective Polyakov Loop actions of the infinite tower of excited states discussed in Sec.~\ref{sec:est}: these are the traces of the Polyakov loop in representations higher than the fundamental one. 
These terms naturally appear in the strong-coupling expansion: they are subleading and are, thus, usually considered as negligible, but we expect that they should become more and more important as the temperature decreases. 
Indeed, it was recently observed~\cite{Bergner:2015rza} that, as the temperature decreases, higher representation terms in the effective action become more and more important, and their contribution is not compensated by an increase in the distance of couplings in the fundamental representation.
At the same time, it is likely that EPL models with long-range interactions with a powerlike decrease of the coupling constants (as those recently proposed in Refs.~\cite{Greensite:2013yd,Greensite:2013bya,Greensite:2014isa,Hollwieser:2016hne}) could be characterized by a much richer spectrum than standard nearest-neighbour models, and they could represent another strategy to obtain larger values of the $\xiratio$ ratio.
We hope that the analysis we propose in this paper could help to fine-tune various proposal for EPL actions and to ensure a correct description of the original lattice gauge theory even at temperatures significantly lower than the critical one.

\vskip1.0cm 
\noindent\textbf{Acknowledgements}\\
%\begin{acknowledgments}
The lattice simulations were run on the GALILEO supercomputer at CINECA. 
We would like to thank Ferdinando Gliozzi and Marco Panero for helpful discussions and comments. 
%\end{acknowledgments}

\bibliography{csi2}

\end{document}